%% file: p133.tex
\documentclass[journal,epsf]{IEEEtran}
\usepackage{amsmath}
\usepackage{amsfonts}
\usepackage{amssymb}
\usepackage{graphicx}

\ifCLASSINFOpdf
\else
\fi
\newcommand {\dfn} {\stackrel{\Delta} {=}}
\newcommand {\exe} {\stackrel{\cdot} {=}}

\newcommand {\bh} {\mbox{\boldmath $h$}}

\newcommand {\bp} {\mbox{\boldmath $p$}}

\newcommand {\hs} {\hat{s}}
\newcommand {\hx} {\hat{x}}
\newcommand {\hbx} {\hat{\mbox{\boldmath $x$}}}

\newcommand {\bx} {\mbox{\boldmath $x$}}
\newcommand {\by} {\mbox{\boldmath $y$}}

\newcommand{\calD}{{\cal D}}
\newcommand{\calE}{{\cal E}}

\newcommand{\calX}{{\cal X}}

\begin{document}
\title{Another Look at the Physics of Large Deviations
With Application to Rate--Distortion Theory}
\author{Neri Merhav
\thanks{N.~Merhav is with the Department
of Electrical Engineering, Technion -- Israel Institute of Technology, Haifa,
32000, Israel. E-mail: merhav@ee.technion.ac.il.}}

\maketitle

\begin{abstract}
\boldmath
We revisit and extend the physical interpretation recently given to a certain identity
between large--deviations rate--functions (as well as applications of this
identity to Information
Theory), as an instance of thermal equilibrium between 
several physical systems that are brought into
contact. Our new interpretation, of mechanical
equilibrium between these systems, is shown to have several advantages
relative to that of thermal equilibrium. This physical point of view also
provides a trigger to the development of certain alternative 
representations of the rate--distortion function and
channel capacity, which are new to the best knowledge of the author.
\end{abstract}
\begin{IEEEkeywords}
Large deviations theory, Chernoff bound, statistical physics, free energy
mechanical equilibrium, rate--distortion theory.
\end{IEEEkeywords}
\IEEEpeerreviewmaketitle

\section{Introduction}

\IEEEPARstart{R}{elationships} between information theory and
statistical physics have been widely recognized in the last few decades,
from a wide spectrum of aspects. These include conceptual aspects, of
parallelisms and analogies 
between theoretical principles in 
the two disciplines, as well as technical aspects, of
mapping between mathematical formalisms in both fields and borrowing
analysis techniques from one field to the other. One example
of such a mapping, is between the paradigm of random codes for channel coding 
and certain models of magnetic materials, 
most notably, Ising models and spin glass 
models (cf.\ e.g., \cite{MM09} and many
references therein).
Today, it is quite widely believed 
that research in the intersection between information theory and statistical
physics may have the potential of fertilizing both disciplines.

This paper is more related to the former aspect mentioned above, namely,
the relationships between the two areas in the conceptual level.
In particular, we revisit results of a recent work \cite{Merhav08},
and propose a somewhat different perspective, which as we believe, has
certain advantages, that will be explained and shown in the sequel.

More specifically, in \cite{Merhav08},
an identity between two forms of the rate function
of a certain large deviations event was established, with several applications
in information theory. Inspired by a few earlier works (cf.\ e.g.,
\cite{McAllester}, \cite{Rose94}, \cite{Shinzato}), this identity was 
interpreted as {\it thermal equilibrium} 
between several many--particle physical systems
that are brought in contact. In particular, the parameter that undergoes
optimization of the Chernoff bound, 
henceforth referred to as the {\it Chernoff
parameter}, plays a role that is intimately related to the equilibrium
temperature: in fact, it is the reciprocal of the temperature, called the
{\it inverse temperature}. The corresponding large deviations rate function
is then identified with the entropy of the system. 

While this physical interpretation 
is fairly reasonable, it turns out, as we show in this
paper, that it leaves quite 
some room for improvement, and we will mention
here just two points. The first, is
that this interpretation does not
generalize to rate functions of combinations of two or more rare events, where
the number of Chernoff parameters is as the number of events. This is because
there is only one temperature parameter in physics.
The other point, which is on a more technical level, is
the following (more details and clarifications will follow 
in Subsection 2B below): while the log--moment generating
function, pertaining to the large deviations 
rate function, naturally includes 
weighting by probabilities, its physical analogue, 
which is the {\it partition
function}, does not. If these probabilities 
are subjected to optimization (e.g., optimization of random coding
distributions), they may depend on the Chernoff parameter,
i.e., on the temperature, in a rather complicated manner, and
then the resulting expression can no longer really be viewed as a
partition function.

In this paper, we propose to interpret the above--mentioned identity of rate
functions as an instance of {\it mechanical equilibrium} (i.e.,
balance between mechanical forces), rather than thermal equilibrium,
and then the Chernoff parameter plays the physical role of an external {\it
force}, or {\it field}, applied to the 
physical system in consideration. In this paradigm,
the large deviations rate function has a natural
interpretation as the (Helmholtz) {\it free energy} of the system, rather than
as entropy. Accordingly, since the rate--distortion function (and similarly,
also channel capacity) can be thought of as a large deviations rate function,
it can also be interpreted as the free energy of a certain system.

This interpretation has several advantages.
First, it is consistent with the analogy between the free energy
in physics and the Kullback--Leibler divergence in information theory (see,
e.g., \cite{Bagci07},\cite{Qian01}), which is well known to play a role as a rate function
when the large deviations analysis is approached by the method of types
\cite{CK81}. Second, it 
is free of the limitations mentioned in the
previous paragraph, as we will see in the sequel.
Third, it serves as a trigger to develop certain
representations of the rate--distortion function (and analogously, the channel
capacity), which are new to the best knowledge of the author.

Since the rate--distortion function
can be thought of as free energy, as mentioned above, one of the 
representations of the rate--distortion function expresses it
as (the minimum achievable) mechanical work carried out by the aforementioned
external force, along a `distance' that is measured in terms of the distortion.
Another representation, which follows from the first one, is as an integral that
involves the single--letter minimum mean square error (MMSE) in estimating the distortion
given the source symbol, according to a certain joint distribution of these
two random variables. The latter representation may suggest
a new route to the derivation of upper and lower bounds on the
rate--distortion function and channel capacity, using the plethora of
upper and lower bounds on MMSE, available from estimation theory.
In particular, for upper bounds, one may examine the
mean squared error of an 
arbitrary estimator, e.g., the best linear estimator. Lower bounds, like the
Bayesian Cram\'er--Rao bound and numerous others are
available in the literature (cf.\ e.g., \cite{WW85},\cite{Weiss85} and references therein).
We have not explored these
directions, however,
in the framework of the work presented herein.

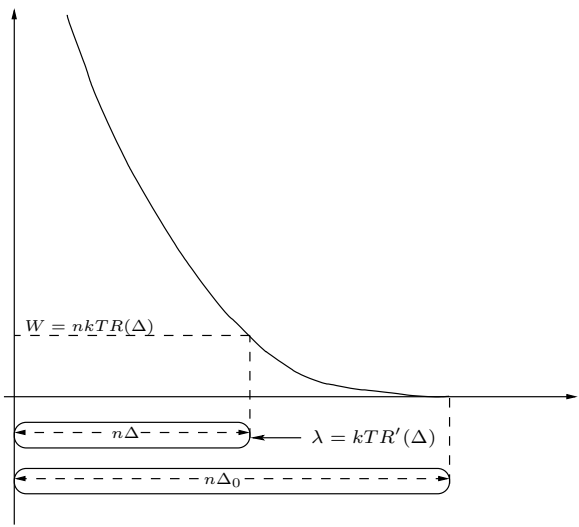
\begin{figure}[ht]
\hspace*{1cm}\input{rd.pstex_t}
\caption{Emulation of $R(\Delta)$ by a physical system.}
\label{rd}
\end{figure}

An additional byproduct of the proposed perspective is the following: Given a source
distribution and a distortion measure, we can describe (at least
conceptually) a concrete physical system that emulates the rate--distortion problem
in the following manner (see Fig.\ \ref{rd}):
When no force is applied to
the system, its total length is $n\Delta_0$,
where $n$ is the number of particles in the system 
(and also the block length in the rate--distortion problem),
and $\Delta_0$ is the distortion
corresponding to zero coding rate.
If one applies to the system a contracting force, that increases
from zero force to some final force 
$\lambda$, such that the length of the system shrinks to 
$n\Delta$, where $\Delta < \Delta_0$ is analogous a prescribed distortion level,
then the following two facts hold true:
(i) An {\it achievable lower bound}
on the total amount of
mechanical work that must be carried out
by the contracting force in order to shrink the system to length $n\Delta$, is
given by
$$W\ge nkTR(\Delta),$$
where $k$ is Boltzmann's constant,
$T$ is the temperature, and $R(\Delta)$ is
the rate--distortion function. (ii) The final force
$\lambda$ is related to $\Delta$ according to
$\lambda=kTR'(\Delta)$, where $R'(\cdot)$ is the derivative of $R(\cdot)$.

Thus, we observe that $R(\Delta)$ plays a role of a fundamental limit, not only in information
theory, but also in physics.

The outline of the paper is as follows. In Section 2, we provide some
background in physics (Subsection 2A) and give a brief description of the physical
interpretation proposed in \cite{Merhav08} (Subsection 2B). Then,
we develop the new proposed physical interpretation, first for a
generic large deviations rate--function (Section 3), and then, in the context of the
rate--distortion problem (Section 4). In Section 5, we present the above mentioned
alternative representations of the rate--distortion function. Finally, in
Section 6, we summarize this work and conclude.

\section{Preliminaries}
\label{bkgd}

\subsection{Physics Background}

Consider a physical system with a large number $n$ of particles,
which can be in a variety of microscopic states (`microstates'),
defined by combinations of, e.g.,
positions, momenta,
angular momenta, spins, etc., of all $n$ particles.
For each such
microstate of the system, which we shall
designate by a vector $\bx=(x_1,\ldots,x_n)$, there is an
associated energy, given by an Hamiltonian (energy function),
$\calE(\bx)$. For example, if $x_i=(\bp_i,h_i)$, where
$\bp_i$ is the momentum vector of particle number $i$ and
$h_i$ is its height, then classically, 
$$\calE(\bx)=\sum_{i=1}^n
\left(\frac{\|\bp_i\|^2}{2m}+mgh_i\right),$$ 
where $m$ is the mass of each particle
and $g$ is the gravitation constant.

One of the most
fundamental results in statistical physics (based on the
law of energy conservation and
the basic postulate that all microstates of
the same energy level are equiprobable)
is that when the system is in thermal
equilibrium with its environment, the probability of a microstate $\bx$ is
given by the {\it Boltzmann--Gibbs} distribution
\begin{equation}
\label{bd}
P(\bx)=\frac{e^{-\beta\calE(\bx)}}{Z_n(\beta)}
\end{equation}
where $\beta=1/(kT)$, $T$ being temperature, $k$ being Boltzmann's constant, and
$Z_n(\beta)$ is the normalization constant,
called the {\it partition function}, which
is given by
$$Z_n(\beta)=\sum_{\bx} e^{-\beta\calE(\bx)}$$
or
$$Z_n(\beta)=\int d\bx e^{-\beta\calE(\bx)},$$
depending on whether $\bx$ is discrete or continuous. The role
of the partition function is by far deeper
than just being a normalization factor, as
it is actually the key quantity from which many
macroscopic physical quantities can be derived,
for example, the Helmholtz free energy\footnote{The physical meaning of
the Helmholtz free energy is the following: The difference between the
Helmholtz free energies of two equilibrium states is
the minimum work that should be done on the system 
in any process of fixed temperature (isothermal process) in the passage between these
two states.
The minimum is obtained when the process is reversible (slow, quasi--static
changes in the system).}
is $-\frac{1}{\beta}\ln Z_n(\beta)$, the average internal
energy (i.e., the expectation of $\calE(\bx)$ where
$\bx$ drawn is according (\ref{bd}))
is given by the negative derivative of $\ln Z_n(\beta)$, the heat capacity
is obtained from
the second derivative, etc. One of the ways to
obtain eq.\ (\ref{bd}), is as the maximum entropy
distribution under an energy constraint
(owing to the second law of thermodynamics), where $\beta$ plays the role of
a Lagrange multiplier that controls this energy level.

Under certain assumptions on the
Hamiltonian function, the following relations are well--known to hold
and can be found in any textbook on elementary statistical physics (see,
e.g., \cite{Beck76},\cite{Kubo61},\cite{MM09}):
Defining the per--particle entropy, $S(E)$, associated with
per--particle energy $E=\calE(\bx)/n$, as $\lim_{n\to\infty}[\ln
\Omega(E)]/n$,\footnote{
Actually, the definition should also include a factor
of $k$, which we will omit in this discussion,
thus considering $S(E)$ as the per--particle entropy in units of $k$.}
(provided that the limit exists), where $\Omega(E)$ is the number of microstates
$\{\bx\}$ with energy level $\calE(\bx)=nE$, then
similarly as in the method of types, one can evaluate $Z_n(\beta)$
defined above, as 
$$Z_n(\beta)=\sum_E \Omega(E)e^{-\beta E}$$
(in the discrete case),
which is of the exponential order
of 
$$\exp\{n\max_E[S(E)-\beta E]\}.$$ 
Defining
$$\phi(\beta)=\lim_{n\to\infty}\frac{\ln Z_n(\beta)}{n},$$
and the Helmholtz free--energy per--particle as 
$$F(\beta)=-\frac{\phi(\beta)}{\beta},$$
we obtain the Legendre relation 
$$\phi(\beta)=\max_E[S(E)-\beta E],$$
where here $E=E(\beta)$ is the maximizer of $[S(E)-\beta E]$.
For a given $\beta$, the Boltzmann--Gibbs
distribution has a sharp peak (for large $n$) at
the level of $E(\beta)$ Joules per--particle.
Assuming that $S(\cdot)$ is concave (which is normally the case),
the above Legendre relation
can be inverted to obtain
$$S(E)=\min_{\beta\ge 0}[\beta E+\phi(\beta)],$$
and both relations can be identified with the thermodynamical
definition of the Helmholtz free energy as 
$$F=E-TS.$$
In the latter relation, the minimizing $\beta=\beta(E)$
(the inverse function of
$E(\beta)$) is the equilibrium inverse temperature associated
with energy level $E$. The second law of thermodynamics asserts
that in an isolated system (which does not exchange energy with
its environment), the total entropy cannot decrease, and hence in
equilibrium, it reaches its maximum. When the system is allowed
to exchange heat with the environment
(at constant volume and temperature), this maximum entropy
principle is replaced by the {\it minimum free energy} principle:
The Helmholtz free energy cannot increase, and it reaches its minimum in equilibrium.

When the Hamiltonian is additive,
that is, 
$$\calE(\bx)=\sum_i\calE(x_i),$$ 
then $P(\bx)$
has a product form (the particles do not interact),
and then the above mentioned physical quantities per particle can be
extracted from the case $n=1$. In this additive case,
the Legendre transform, that takes $\phi(\beta)$ to $S(E)$,
is similar to the Legendre transform that defines the 
rate function (the exponent of the Chernoff bound)
pertaining to the probability of the event 
$$\sum_{i=1}^n\calE(x_i)\le nE,$$
thus the parameter to be optimized in the Chernoff bound plays the role
of inverse temperature in the corresponding statistical--mechanical system.

Another look at this correspondence between large deviations rate functions
and thermal equilibrium is the following: If $P$ is the above mentioned
Boltzmann--Gibbs distribution and $Q$ is another probability distribution
on the micorstates $\{\bx\}$, then, as is shown e.g., in \cite{Bagci07},
the Kullback--Leibler divergence between $Q$ and $P$ is given by
$$D(Q\|P)=\beta(F_Q-F_P),$$ 
where $F_P$ and $F_Q$ are, respectively, the
Helmholtz free energies pertaining to $P$ and $Q$. The rate function pertaining to a
large deviations event is normally given by the minimum divergence under the
constraints corresponding to this event (see, e.g., \cite[Chap.\ 11]{CT06}),
and so, it is equivalent to minimum free energy, i.e., thermal equilibrium by
the second law.

Consider next a system of $n$ non--interacting particles as before, except
that now the Hamiltonian
is shifted by a quantity that is proportional to some parameter $\lambda$, i.e.,
the Hamiltonian is redefined as
$$\calE(\bx,\by) = \calE_0(\bx)-\lambda\cdot\sum_{i=1}^n y_i,$$
where we have changed the notation of the (original) Hamiltonian to
$\calE_0(\bx)$, and
where $\{y_i\}$ are some additional variables used to describe 
the microstate. These new variables may 
either be dependent or independent 
of the original microstate variables $\{x_i\}$ 
(both cases are demonstrated in Example 1 below)
and their number, $n$, is here taken to be the same as the number of
$\{x_i\}$, primarily, for reasons of convenience.\footnote{In general, their number can
be different, but then it is still assumed to grow proportionally to $n$.}
The parameter $\lambda$ is thought of as an external control parameter, i.e.,
a {\it driving force} (or a field) that acts on the system via
the state variables $\{y_i\}$.
The parameter $\lambda$ 
can be a mechanical force (e.g., pressure, elastic extraction/contraction force, gravitational
force), an electric field (acting on an a charged particle or
an electric dipole), 
a magnetic field (acting on a magnet or spin), or even a chemical
driving force (chemical potential).\\

\noindent
{\it Example~1} (may be skipped without loss of continuity).
Consider the following two systems. The first 
is the same example as in the first paragraph of this subsection, 
namely, non--interacting particles in motion under gravitation. 
The Hamiltonian, 
$$\sum_i\left(\frac{\|\bp_i\|^2}{2m}+mgh_i\right)$$ 
can be thought of as
being composed of the `original' Hamiltonian $\sum_i\|\bp_i\|^2/(2m)$
(with $\{\bp_i\}$ replacing $\{\bx_i\}$), and the `shifting' term,
$mg\sum_ih_i$, whose force parameter is $\lambda=-mg$ (gravitational force),
acting on the height variables $y_i=h_i$.
In this example, the variables $\bx=\bp$ and $\by=\bh=(h_1,\ldots,h_n)$
are independent. The second system consists of $n$ one--dimensional harmonic
oscillators (e.g., springs or pendulums),
where the Hamiltonian is 
$$\sum_i\left(\frac{\|p_i\|^2}{2m}+\frac{Ky_i^2}{2}\right),$$
$p_i$ being the (one--dimensional) momentum, $y_i$ -- the displacement of
each oscillator from its equilibrium position, and
$K$ is the elasticity constant. Now, suppose that an external force $\lambda$
is applied to each spring, so the Hamiltonian becomes
$$\sum_i\left(\frac{\|p_i\|^2}{2m}+\frac{Ky_i^2}{2}-\lambda y_i\right).$$ 
In this case, 
the variables of the original Hamiltonian $x_i=(p_i,y_i)$ contain the
variables $\{y_i\}$, of the shifting term, as a subset. 
We also see that the modified Hamiltonian is, within
an immaterial additive 
constant, identical to
$$\sum_i\left[\frac{\|p_i\|^2}{2m}+\frac{K}{2}\left(y_i-\frac{\lambda}{K}\right)^2\right].$$
This means that the
force $\lambda$ shifts the common mean of the RV's $\{y_i\}$, which is 
equilibrium point of all oscillators, by $\Delta y=\lambda/K$, as
expected. This concludes Example 1. $\Box$

Consider next the partition function
$$\tilde{Z}_n(\beta,\lambda)=\sum_{\bx,\by}e^{-\beta[\calE_0(\bx)-\lambda\sum_iy_i]}.$$
The {\it Gibbs free energy}\footnote{The Gibbs free energy
has a meaning similar to the of the Helmholtz free energy (see footnote no.\
1), but
it refers to partial work: the difference
between the Gibbs free energies of two equilibrium points is the minimum amount
of work to be done on the system,
{\it other than work pertaining to changes in the variables} $\{y_i\}$, 
in an isothermal process with
fixed $\lambda$, in the passage between these two points.}
per particle is defined as
$$G_n(\beta,\lambda)= -\frac{kT\ln \tilde{Z}_n(\beta,\lambda)}{n}$$
and the asymptotic Gibbs free energy per particle is
$$G(\beta,\lambda)=\lim_{n\to\infty}G_n(\beta,\lambda).$$
What is the relation between between the Helmholtz free energy
and the Gibbs free energy?
Let $\Omega(E,Y)\sim e^{nS(E,Y)}$ denote the number of 
microstates $\{(\bx,\by)\}$ for which 
$$\sum_i\calE_0(x_i)=nE~~\mbox{and}~~\sum_iy_i=nY.$$
Then, defining the partial partition function
$$Z_n(\beta,Y)=\sum_{\{(\bx,\by):~\sum_iy_i=nY\}}e^{-\beta\calE_0(\bx)},$$
the normalized Helmholtz free energy for a given $Y$
$$F_n(\beta,Y)=-\frac{kT\ln Z_n(\beta,Y)}{n},$$ 
and the corresponding asymptotic
normalized Helmholtz free energy, 
$$F(\beta,Y)=\lim_{n\to\infty}F_n(\beta,Y),$$ 
we have (similarly as in the method of
types):
\begin{eqnarray}
e^{-\beta n
G_n(\beta,\lambda)}&=&\sum_{\bx,\by}
e^{-\beta[\calE_0(\bx)-\lambda\sum_iy_i]}\nonumber\\
&=&\sum_{E,Y}\Omega(E,Y)e^{-\beta(nE-\lambda nY)}\nonumber\\
&\exe&\sum_{E,Y}e^{n[S(E,Y)-\beta(E-\lambda Y)]}\nonumber\\
&=&\sum_{Y}e^{n\beta\lambda Y}\sum_E e^{n[S(E,Y)-\beta E]}\nonumber\\
&=&\sum_{Y}e^{n\beta\lambda Y}Z_n(\beta,Y)\nonumber\\
&=&\sum_{Y}e^{n\beta\lambda Y}\cdot e^{-\beta nF_n(\beta,Y)}\nonumber\\
&\exe&\exp\{n\beta\cdot\max_Y[\lambda Y-F(\beta,Y)\}
\end{eqnarray}
where $\exe$ denotes asymptotic equivalence in the exponential
scale.\footnote{More 
precisely, $a_n\exe b_n$, for two positive sequences $\{a_n\}$ and
$\{b_n\}$, means that $\frac{1}{n}\log\frac{a_n}{b_n}\to 0$, as $n\to\infty$.}
This results in the Legendre relation
$$G(\beta,\lambda)=\min_Y[F(\beta,Y)-\lambda Y].$$
Assuming that $F(\beta,Y)$ is convex in $Y$ for fixed $\beta$,
the inverse Legendre relation is
\begin{eqnarray}
\label{FLegendre}
F(\beta,Y)
&=&\max_{\lambda}[G(\beta,\lambda)+\lambda Y]\nonumber\\
&=&\max_{\lambda}\left[\lambda Y-kT\times\right.\nonumber\\
& &\left.\lim_{n\to\infty}\frac{1}{n}\ln\left(\sum_{\bx,\by}
e^{-\beta[\calE_0(\bx)-\lambda\sum_iy_i]}\right)\right]\nonumber\\
&=&kT\cdot\max_{\lambda}\left[\beta\lambda
Y-\right.\nonumber\\
& &\left.\lim_{n\to\infty}\frac{1}{n}\ln\left(\sum_{\bx,\by}
e^{-\beta\calE_0(\bx)}\cdot e^{\beta\lambda\sum_iy_i}\right)\right]\nonumber\\
&=&kT\cdot\max_{s}\left[sY-\right.\nonumber\\
& &\left.\lim_{n\to\infty}\frac{1}{n}\ln\left(\sum_{\bx,\by}
e^{-\beta\calE_0(\bx)}\cdot e^{s\sum_iy_i}\right)\right]
\end{eqnarray}
where in the last step, we changed the optimization variable $\lambda$ to
$s=\beta\lambda$ for fixed $\beta$. Since $s$ is proportional to $\lambda$
for fixed $\beta$, and $\lambda$ designates force, 
we will henceforth refer to $s$ also as `force' (although its physical
units are different). We will get back to eq.\ (\ref{FLegendre}) soon.

\subsection{A Brief Summary of \cite{Merhav08}}

First, recall that in the previous subsection, 
we mentioned that the Legendre relation 
$$S(E)=\min_{\beta\ge 0}[\beta E
+\phi(\beta)]$$ 
is similar to the rate function of the large deviations
event $\{\sum_i\calE(x_i)\le nE\}$ for i.i.d.\ RV's $\{x_i\}$, governed by a given
distribution $P$. The difference is that in the latter, 
the log--moment generating function
$$\ln\sum_xP(x)e^{-\beta\calE(x)},$$
that undergoes the Legendre transform, 
contains weighting by the probabilities $\{P(x)\}$, unlike
the log--partition 
$$\ln\sum_xe^{-\beta\calE(x)},$$ 
which does not.
In \cite{Merhav08} it was proposed to interpret the weights $\{P(x)\}$ as
being proportional to a factor of the multiplicity of states $\{x\}$ having the same
energy $\calE(x)$, i.e., as the {\it degeneracy} in the physics
terminology.\footnote{Another approach, proposed in \cite{SK08}, was to
absorb $P(x)$ as part of the Hamiltonian, but then the Hamiltonian becomes
temperature--dependent, but this does not comply with the common paradigm in statistical
mechanics.} 

When considering applications of large deviations theory
to information theory, one can view the rate--distortion function (and
analogously, also channel capacity) as the large--deviations rate function of
the event $\{\sum_{i=1}^nd(x_i,\hat{x}_i)\le n\Delta\}$, where
$\bx=(x_1,\ldots,x_n)$ is a given typical source sequence (i.e., its empirical
distribution agrees with the source $P$) and $\{\hx_i\}$ are i.i.d.\ RV's
drawn by a certain random coding distribution $Q$. As was observed in
\cite{Merhav08}, there are two ways to express the large deviations
rate function of this event, which is also the rate--distortion
function, $R_Q(\Delta)$, for the given random distribution $Q$: 
The first is by considering all distortion variables
$\{d(x_i,\hx_i)\}$ together, on the same footing, resulting in
the expression
$$I(\Delta)=-\min_{\beta\ge 0}\left[\beta\Delta+\sum_x
P(x)\ln\sum_{\hx}Q(\hx)e^{-\beta d(x,\hx)}\right],$$
which can also be obtained (see, e.g., \cite[p.\ 90, Corollary 4.2.3]{Gray90}) using different
considerations. The second way is to 
separate the distortion contributions,
$\{\Delta_x\}$, allocated 
to the various source letters $\{x\}$, which results in
\begin{eqnarray}
I(\Delta)&=&-\max_{\{\Delta_x\}:~\sum_xP(x)\Delta_x\le
\Delta}\sum_xP(x)\min_{\beta_x\ge 0}\left[\beta\Delta_x+\right.\nonumber\\
& &\left.\ln\sum_{\hx}Q(\hx)e^{-\beta_x d(x,\hx)}\right].\nonumber
\end{eqnarray}
The identity between these two expressions, as was proved in \cite{Merhav08},
means that the outer maximum in the second expression (maximum
entropy) is achieved when $\{\Delta_x\}$ are allocated in such a way that
the minimizing temperature parameters $\{\beta_x\}$ are all the same, namely,
thermal equilibrium between all subsystems indexed by $x$. Once again,
$\{Q(\hx)\}$ can be interpreted as degeneracy, which is fine as long as $Q$ is
fixed. However, the real rate--distortion function, 
$R(\Delta)=\min_QR_Q(\Delta)$,
is obtained by optimization (of either expression) over $Q$ and the optimum $Q$
may, in general, depend on $\beta$ (or equivalently, on $\Delta$).
In this situation, $Q$ can no longer be given the meaning
of degeneracy, because in physics, degeneracy has nothing to do with
temperature.

Another limitation of interpreting $\beta$ as temperature, is that it does not
extend to two or more rare events at the same time. For instance, the
rate--distortion
function $R_Q(\Delta_1,\Delta_2)$ w.r.t.\ two simultaneous
distortion constraints, with distortion measures $d_1$ and $d_2$, is given by
the two--dimensional Legendre transform
\begin{eqnarray}
R_Q(\Delta_1,\Delta_2)
&=&-\min_{\beta_1\le 0}\min_{\beta_2\le 0}\left[\beta_1\Delta_1+\beta_2\Delta_2+
\sum_{x\in\calX}P(x)\times\right.\nonumber\\
& &\left.\ln\left(
\sum_{\hx}Q(\hx)e^{-\beta_1d_1(x,\hx)-\beta_2d_2(x,\hx)}\right)\right].
\end{eqnarray}
But this does not have any apparent physical interpretation because there is
only one temperature in physics.

\section{Large Deviations and Free Energy}

The main idea in this paper is 
that in order to give a physical interpretation to the rate function as the
Legendre transform of the log--moment generating function, we use 
the Legendre transform that relates the
Helmholtz free energy to the Gibbs free energy, 
$G(\beta,\lambda)$ (cf.\ eq.\ (\ref{FLegendre})), rather than the one
that relates the Helmholtz free 
energy to the entropy, $S(E)$. Thus, the Chernoff
variable would be the force $\lambda$ (or $s$) rather than the inverse temperature
$\beta$. Also, considering the temperature as being fixed throughout, we
can view the weights $\{Q(\hx)\}$ (in the rate--distortion application) as
part of the Hamiltonian $\calE_0$, 
which now may depend on the control parameter
$\lambda$. This also allows combinations of two or more large deviations events
since one may consider a system that is subjected to more than one force,
e.g., two or three components of same force, 
or a superposition of different types of
forces.

Specifically, let us first
compare the Helmholtz free energy expression (\ref{FLegendre})
to the rate function \cite{DZ93} of the simple large deviations
event $\{\sum_iy_i \ge nY\}$ w.r.t.\ some probability distribution $P$:
$$I(Y)=\max_s
\left[sY-\lim_{n\to\infty}\frac{1}{n}\ln\left(\sum_{y}P(\by)e^{s\sum_iy}\right)\right]$$
which in the case where $\{y_i\}$ are i.i.d.\ ($P(\by)=\prod_iP(y_i)$), boils
down to
$$\max_s
\left[sY-\ln\sum_{y}P(y)e^{sy}\right].$$
Fixing the temperature $T$ to some $T_0=1/(k\beta_0)$, taking $\by\equiv\bx$ and
$\calE_0(\bx)\equiv\calE_0(\by)=-kT_0\ln
P(\by)$, we readily see that
$I(Y)$ coincides with $F(\beta_0,Y)$ up to the multiplicative constant factor of
$kT_0$, which is immaterial.
We observe then that the
large deviations rate function has a natural
interpretation as the Helmholtz free energy (in units of $kT_0$) of a system with
Hamiltonian $$\calE_0(\by)=-kT_0\ln P(\by)$$ 
and temperature $T_0$.
As said, the Chernoff parameter $s$ has 
(again, within the factor $\beta_0$) the meaning of a driving force that acts on the
displacement variables $\{y_i\}$ (cf.\ e.g.,
the above example of the one--dimensional
harmonic oscillator, which makes it explicit). For example, in 
the i.i.d.\ case, the driving force $s$ required to shift the expectation
of each $y_i$ (and hence also 
of $\frac{1}{n}\sum_i y_i$) towards $Y$, which is the solution to the equation
$$Y=\frac{\partial}{\partial s}\ln \sum_y P(y)e^{sy}$$
or equivalently,
$$Y=\frac{\sum_y P(y)\cdot ye^{sy}}{\sum_y P(y)\cdot e^{sy}}.$$
The Legendre transform relation between
the log--partition function and $I(Y)$ induces a one--to--one mapping between $Y$ and $s$
which is defined by the above equation. To emphasize this dependency, we
henceforth denote the value of $Y$, corresponding to a given $s$, by
$\left<y\right>_s$, which symbolizes the fact that it is
the expectation\footnote{
In the sequel, we use $\left<\cdot\right>_s$ to denote other moments of $y$
w.r.t.\ $P_s$ as well.}
of each $y_i$, denoted generically by $y$, w.r.t.\ the
probability distribution $P_s=\{P_s(y)\}$, where 
$$P_s(y)=P(y)e^{sy}/[\sum_{y'}P(y')e^{sy'}],$$
i.e.,
$$\left<y\right>_s=\frac{\sum_y P(y)\cdot ye^{sy}}{\sum_y P(y)\cdot e^{sy}}=
\frac{\partial}{\partial s}\ln \sum_y P(y)e^{sy}.$$
On substituting
$\left<y\right>_s$ instead of $Y$ in the
expression defining $I(Y)$, we can re-define the rate function as a function of
(the maximizing) $s$, i.e.,
$$\hat{I}(s)=s\left<y\right>_s-\ln\sum_y
P(y)e^{sy}.$$
Note that $\hat{I}(s)$ can be represented in an integral form as follows:
\begin{eqnarray}
\hat{I}(s)&=&\int_0^{s}
\mbox{d}\hs\cdot\left(\left<y\right>_{\hs}+
\hs\frac{\mbox{d}\left<y\right>_{\hs}}{\mbox{d}\hs}-\left<y\right>_{\hs}\right)\nonumber\\
&=&\int_{\left<y\right>_0}^{\left<y\right>_{s}}\hs\cdot\mbox{d}\left<y\right>_{\hs}.
\end{eqnarray}
Now observe that the integrand 
is a product of the force, $\hs$, and an infinitesimal
displacement that it works upon,
$\mbox{d}\left<y\right>_{\hs}=\left<y\right>_{\hs}-\left<y\right>_{\hs-d\hs}$ 
(which in turn is the response of the system to
a corresponding infinitesimal change in the force from $\hs-\mbox{d}\hs$ to
$\hs$).
In physical terms, $\hs\cdot\mbox{d}\left<y\right>_{\hs}$
is therefore an infinitesimal contribution of the average {\it work} (in units
of $kT_0$) done by
the driving force $\hs$ on the displacement variables $\{y_i\}$.
Thus, the integral, $\hat{I}(s)=\int 
\hs\cdot\mbox{d}\left<y\right>_{\hs}$ is the total
amount of work (again, in units of $kT_0$) 
carried out by the force $\hs$, as it increases from
zero to $s$ during a slow process that allows the system to equilibrate after
every infinitesimally small change in $\hs$. In the language of physics,
this is a {\it reversible process}, or a {\it quasi-static process}. Using the
concavity of $F$ as a function of $s$, it is easy to show that
any protocol of changing $\hs$ from $0$ to $s$, in a way that includes abrupt
changes in $\hs$,
would always yield an amount of work larger than or equal to $\hat{I}(s)$
(which is
consistent with the operative meaning of $\hat{I}(s)$ as the free energy of
the system -- see footnote no.\ 1). Thus, for any sequence, $s_1,\ldots,
s_\ell$, of numbers between $0$ and $s$, we can sandwich $\hat{I}(s)$ between two bounds
$$\sum_{i=1}^{\ell-1}s_i(\left<y\right>_{s_{i+1}}-\left<y\right>_{s_i})
\le \hat{I}(s)\le
\sum_{i=1}^{\ell-1}s_{i+1}(\left<y\right>_{s_{i+1}}-\left<y\right>_{s_i}),$$
which become tighter and tighter as the partition of the interval $[0,s]$,
defined by $\{s_i\}_{i=1}^{\ell}$,
becomes more refined.

For an alternative integral expression, one
observes that
$\mbox{d}\left<y\right>_s/\mbox{d}s=\left<y^2\right>_s-\left<y\right>_s^2\dfn\mbox{Var}_s\{y\}$, 
namely, the variance of $y$
w.r.t.\ the probability distribution $P_s$. Thus,
$$\hat{I}(s)=\int_0^s\hs\cdot\mbox{Var}_{\hs}\{y\}\mbox{d}\hs$$
and
$$\left<y\right>_s=\left<y\right>_0+\int_0^s\mbox{Var}_{\hs}\{y\}\mbox{d}\hs.$$
Note that, by the same token, 
in the interpretation of \cite{Merhav08}, where the
Chernoff parameter was the inverse temperature $\beta$, that is conjugate
to the Hamiltonian $\calE$, the corresponding
integral could have been represented as
$\int\hat{\beta}\cdot\mbox{d}\left<\calE\right>_{\hat{\beta}}=\int\frac{\mbox{d}Q}{kT}$,
$Q$ being heat, which is the change of entropy along a reversible process.
The corresponding variance expressions would
then be related to the heat capacity at constant volume.
In the more general context considered here, this is a special case
of the fluctuation--dissipation theorem in statistical physics
(cf.\ e.g., \cite[p.\ 32, eq.\ (2.44)]{MM09}). 

We next discuss a physical
example which will be directly relevant for the rate--distortion problem.

\vspace{0.1cm}

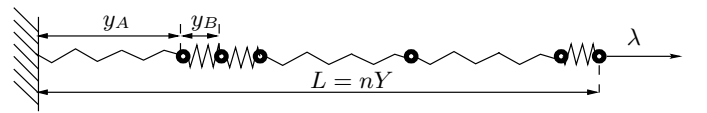
\begin{figure}[ht]
\hspace*{0cm}\input{chain3.pstex_t}
\caption{One--dimensional array of two--state elements.}
\label{chain}
\end{figure}

\noindent
{\it Example~2} \cite[p.\ 134, Problem 13]{Kubo61}: 
Consider a physical system, modeled as a one--dimensional array of 
$n$ elements (depicted as small springs in Fig.\ \ref{chain}), that are
arranged along a straight line.
Each element may
independently be 
in one of two states, $A$ or $B$ (e.g., in state $A$ the element is stretched
and in state $B$, it is
contracted, according to Fig.\ \ref{chain}). 
The state of the
$i$--th element, $i=1,2,\ldots,n$, 
is labeled $\hx_i\in\{A,B\}$. When an element is at state $\hx$, its
length is $y_{\hx}$ and its
internal energy is
$\epsilon_{\hx}$. 
A stretching 
force $\lambda > 0$ (or a contracting force, if $\lambda < 0$) 
is applied to one edge of the array, whereas the other edge
is fixed to a wall. What is the expected (and most
probable) total length $L=nY$ of the array at temperature $T_0$? 

Since the elements are
independent, 
\begin{eqnarray}
& &\tilde{Z}_n(\beta_0,\lambda)\nonumber\\
&=&
\sum_{\hx_1=0}^1\ldots\sum_{\hx_n=0}^1
\exp\left\{-\beta_0\left[\sum_i\epsilon_{\hx_i}-
\lambda\sum_iy_{\hx_i}\right]\right\}\nonumber\\
&=&[e^{-\beta_0(\epsilon_A-\lambda
y_A)}+e^{-\beta_0(\epsilon_B-\lambda y_B)}]^n,
\end{eqnarray}
and so, 
\begin{eqnarray}
nG_n(\beta_0,\lambda)&=&-kT_0\ln
\tilde{Z}_n(\beta_0,\lambda)\nonumber\\
&=&-nkT_0\ln\left[e^{-\beta_0(\epsilon_A-\lambda
y_A)}+e^{-\beta_0(\epsilon_B-\lambda y_B)}\right].\nonumber
\end{eqnarray}
The expected length is
\begin{eqnarray}
nY&=&-n\cdot\frac{\partial G_n(\beta_0,\lambda)}{\partial\lambda}\nonumber\\
&=&
\frac{n[y_A e^{-\beta_0(\epsilon_A-\lambda y_A)}+
y_B e^{-\beta_0(\epsilon_B-\lambda y_B)}]}
{e^{-\beta_0(\epsilon_A-\lambda y_A)}+
e^{-\beta_0(\epsilon_B-\lambda y_B)}}.
\end{eqnarray}
In terms of the foregoing discussion, $s=\beta_0\lambda$, 
the force scaled by $\beta_0$, controls the expected length
per element which is
$$Y=\left<y\right>_s=
\frac{y_A e^{-\beta_0\epsilon_A+sy_A}+
y_B e^{-\beta_0\epsilon_B+sy_B}}
{e^{-\beta_0\epsilon_A+sy_A}+
e^{-\beta_0\epsilon_B+sy_B}}.$$
The free energy per element is then
$$F(\beta_0,Y)=-kT_0\ln\left[e^{-\beta_0\epsilon_A+s
y_A}+e^{-\beta_0\epsilon_B+sy_B}\right]+kT_0sY$$
where $s$ is related to $Y$ according to second to the last equation,
which is also the value of $s$ that maximizes the last expression.

Consider now two arrays as above, labeled by $x\in\{a,b\}$, which
consist of two different types of elements.
Array $x$ has $n(x)$ elements, and as before,
each element of this array 
may be in one of two states, $A$ or $B$. 
When an element of array $x$ is at
state $\hx$, its length is $y_{\hx|x}$ and its internal energy 
is $\epsilon_{\hx|x}$. 
The two arrays are connected together to 
form a larger system 
with a total of $n=n(a)+n(b)$ elements, and this larger system is
stretched (or shrinked) so that its edges are fixed at two points which are
at distance $nY_0$ far apart.
What is the contribution of each individual array to the total length, $nY$,
and what is the force `felt' by each one of them?

Denoting $p_a=n(a)/n$ and $p_b=n(b)/n$,
the total free energy per element is given by 
\begin{eqnarray}
& &p_aF_a(\beta_0,Y_a)+
p_bF_b(\beta_0,Y_b)\nonumber\\
&=&p_aF_a(\beta_0,Y_b)+
p_bF_b\left(\beta_0,\frac{Y_0-p_aY_a}{p_b}\right),
\end{eqnarray}
where $F_a$ and $F_b$ are the Helmholtz free energies per element (cf.\ above) pertaining
to the two arrays, respectively, and $Y_a$ and $Y_b$ are their normalized
lengths.
At equilibrium, $p_a$ minimizes this expression,
and the minimizing $p_a$ solves the equation:
$$\frac{\partial F_a(\beta_0,Y)}{\partial Y}\bigg|_{Y=Y_a}=
\frac{\partial F_b(\beta_0,Y)}{\partial Y}\bigg|_{Y=(Y_0-p_aY_a)/p_b}.$$
But the left--hand side is $\lambda_a=kT_0s_a$, 
the force felt by array (a), and
similarly, the
right--hand side is $\lambda_b=kT_0s_b$, the force felt by array 
(b). The last
equation tells us that in mechanical equilibrium they are equal, which makes
sense, as otherwise the boundary point between the two arrays
would keep moving in either
direction.\footnote{This is similar to the classical mechanical equilibrium between two volumes of gas
separated by a freely moving plate, which stabilizes at the point where the
pressures from both sides equalize.} In other words, the equilibrium values of
$Y_a$ and $Y_b$ are adjusted in a way that
$$F_a(\beta_0,Y_a)=\max_\lambda[G_a(\beta_0,\lambda)+\lambda Y_a]$$
and
$$F_b(\beta_0,Y_b)=\max_\lambda[G_b(\beta_0,\lambda)+\lambda Y_b]$$
would be both maximized by the {\it same} value of $\lambda$ (or,
equivalently, $s$). In this situation, the
same value of $\lambda$ would also achieve the maximum of the weighted sum:
$$\max_\lambda[p_a
G_a(\beta_0,\lambda)+p_bG_b(\beta_0,\lambda)+
\lambda Y_0],$$
which treats the entire system as a whole. The maximizing
value of $\lambda$ is the one that corresponds to total length $Y_0$.
This concludes Example~2. $\Box$

In the next section, we will see how Example 2 (especially, it second part,
with two connected arrays of elements) is directly applicable to the
rate--distortion setting.

\section{Rate--Distortion}

Let us consider now the rate--distortion coding problem.
We are given a source
sequence $\bx=(x_1,\ldots,x_n)$ to be compressed, whose letters $\{x_i\}$ take on values
in a finite alphabet $\calX$ of size $K$. We assume that the source has a
given empirical distribution $P=\{P(x),~x\in\cal X\}$ (typically, close to the
real distribution), i.e., each letter $x\in\calX$ appears $n(x)=nP(x)$ times
in $\bx$. Next consider a random selection of a reproduction codeword
$\hbx=(\hx_1,\ldots,\hx_n)$, where each reproduction symbol $\hx_i$
is drawn i.i.d.\ from a distribution $Q=\{Q(\hx),~\hx\in\hat{\calX}\}$,
where $\hat{\calX}$ is a finite reproduction alphabet of size $J$. 
For the most part of our discussion, it will be assumed 
that even if the desired distortion level varies, the random coding
distribution $Q$ is nevertheless kept fixed, 
for the sake of simplicity.\footnote{A word of clarification is in order
here: Earlier, we mentioned that
the optimum $Q$ may depend on $s$, or equivalently on $\Delta$. 
In the sequel, we describe certain
processes along which the distortion level varies, starting from a very high
distortion level $\Delta_0$, and ending at a given, desired distortion level, $\Delta$. To make a
statement concerning the rate--distortion function, computed at the latter
distortion level, $R(\Delta)$, we can always pick the optimum $Q$ for this
target value of
$\Delta$ and keep it fixed, even when considering the above--mentioned higher distortion
levels. Thus, in these processes, for distortion levels above $\Delta$, we will, in general, 
`move' along the curve $R_Q(\cdot)$, which is the rate--distortion function
with an output distribution constrained to $Q$, rather than the curve
$R(\cdot)$. Of course, the two curves intersect at distortion $\Delta$.
The analysis can be modified to allow $Q$ depend on $s$ along the process (see
comment no.\ 4 on this in Section 6).}
It is well known that the rate--distortion function of the source $P$, w.r.t.\
a given distortion measure $d(x,\hx)$, is given by the rate function
of the large deviations event $\{\sum_{i=1}^nd(x_i,\hx_i)\le n\Delta\}$.

Occasionally, instead of working with the reproduction symbols as our RV's, we
will sometimes work
directly with the
distortions $\{d(x_i,\hx_i)\}$ incurred, which will be denoted by $\{\delta_i\}$
(playing the same role as $\{y_i\}$ thus far).
Accordingly, we define
$$Q(\delta|x)=\sum_{\{\hx:~d(x,\hx)=\delta\}} Q(\hx).$$
Thus, we think of the distortion $\delta$ as a RV drawn from a distribution
$Q(\delta|x)$ indexed by the corresponding source symbol $x$, rather than as a function of $x$ and a RV $\hx$,
whose distribution $Q(\hx)$ does not depend on $x$. 
The large deviations
event under consideration is then $\{\sum_{i=1}^n \delta_i \le n\Delta\}$,
where $\{\delta_i\}$
are still independent, but no longer identically distributed. For each
$x\in\calX$, $n(x)=nP(x)$ of these RV's are drawn from $Q(\delta|x)$.
The large deviations rate function, obtained when all $\{\delta_i\}$ are handled as
a whole, is given by
$$I(\Delta)=\max_s\left[
s\Delta-\sum_{x\in\calX}P(x)\ln\left(\sum_{\delta}Q(\delta|x)e^{s\delta}
\right)\right].$$
In analogy to the results of \cite{Merhav08} (see also Subsection 2A), 
another look is the following: Consider the partial distortions, sorted
according to the underlying source symbols, i.e., for each $x\in\calX$,
$\sum_{i:~x_i=x}\delta_i$ is the total distortion contributed by $x$. Clearly, the large
deviations event under discussion occurs iff there exists a distortion
allocation $\calD=\{\Delta_x,~x\in\calX\}$ with $\sum_{x\in\calX}P(x)\Delta_x\le
\Delta$ such that
$\sum_{i:~x_i=x}\delta_i\le n(x)\Delta_x$ for all $x\in\calX$. Thus, it can be thought
of as the union (over all possible distortion allocations)
of the intersections (over $\calX$) of the independent events
$\{\sum_{i:~x_i=x}y_i\le n(x)\Delta_x\}$. As shown in \cite{Merhav08}, 
since the effective number of distortion
allocations is polynomial in $n$, the probability is dominated by the worst
allocation, which yields
\begin{eqnarray}
\tilde{I}(\Delta)&=&\min_{\{\calD:~\sum_{x\in\calX}P(x)\Delta_x\le
\Delta\}}\sum_{x\in\calX}P(x)\times\nonumber\\
& &\max_{s_x}\left[s_x\Delta_x-
\ln\left(\sum_{\delta}Q(\delta|x)e^{s_x\delta}\right)\right].
\end{eqnarray}
We argue that $\tilde{I}(\Delta)=I(\Delta)$ and hence both coincide with the rate--distortion
function $R_Q(\Delta)$ w.r.t.\ the random coding distribution $Q$. 

Before we prove it formally, we comment that the intuition
comes from interpreting the expressions of the rate functions in the framework
of the above example of stretching/contracting concatenated 
one dimensional arrays of elements. Here, we have $|\calX|=K$
different arrays at temperature $T_0$, concatenated
together to form one larger system with a total of 
$n$ elements. Each individual array is labeled by
$x\in\calX$ and it contains $n(x)=nP(x)$ elements. Each such
element may be in one of $J$ states, labeled by $\hx\in\hat{\calX}$.
The `length' and the internal energy of an element of array $x$ at state $\hx$
are $\delta_{\hx|x}=d(x,\hx)$ and
$\epsilon_{\hx|x}=-kT_0\ln Q(\hx)$ (independent of $x$), respectively. 
Upon identifying this mapping between the rate---distortion problem and
the physical example, we immediately see that their mathematical formalisms, and
hence also their properties, are precisely the same. Indeed,
the expression of $I(\Delta)$ is the Helmholtz
free energy (in units of $kT_0$)
per element (pertaining to the entire system as a
whole) when the total length is shrinked to $n\Delta$. On the other hand, the
expression of $\tilde{I}(\Delta)$ describes the {\it minimum} Helmholtz free energy 
(again, in units of $kT_0$) across all
partial length allocations $\{n(x)\Delta_x\}_{x\in\calX}$ that comply with a total length not
exceeding $n\Delta$. But this minimum free energy is achieved when all
individual arrays `feel'
the same force, i.e., the same value of $s_x$. Hence, the two expressions
should coincide. This means, among other things, that the typical 
relative contribution of each source symbol $x$
to the distortion behaves exactly like the 
relative lengths of the individual arrays
when they lie in mechanical equilibrium. 

Formally, the following proof is similar to 
that of \cite[Theorem 1]{Merhav08}, but for completeness,
we provide it here too. We first prove that $\tilde{I}(\Delta)\ge I(\Delta)$
and then the reversed inequality.
\begin{eqnarray}
\tilde{I}(\Delta)
&=&\min_{\{\calD:~\sum_{x\in\calX}P(x)\Delta_x\le
\Delta\}}\sum_{x\in\calX}P(x)\cdot\max_{s_x\le
0}\left[s_x\Delta_x\right.\nonumber\\
&
&\left.-\ln\left(\sum_{\delta}Q(\delta|x)e^{s_x\delta}\right)\right]\nonumber\\
&=&\min_{\{\calD:~\sum_{x\in\calX}P(x)\Delta_x\le
\Delta\}}\sum_{x\in\calX}\max_{s_x\le
0}\left[s_xP(x)\Delta_x-\right.\nonumber\\
&
&\left.P(x)\ln\left(\sum_{\delta}Q(\delta|x)e^{s_x\delta}\right)\right]\nonumber\\
&\ge&\min_{\{\calD:~\sum_{x\in\calX}P(x)\Delta_x\le
\Delta\}}\max_{s\le 0}\sum_{x\in\calX}\left[sP(x)\Delta_x-\right.\nonumber\\
&&\left.P(x)\ln\left(\sum_{\delta}Q(\delta|x)e^{s\delta}\right)\right]\nonumber\\
&\ge&\min_{\{\calD:~\sum_{x\in\calX}P(x)\Delta_x\le
\Delta\}}\max_{s\le 0}
\left[s\sum_{x\in\calX}\Delta_xP(x)-\right.\nonumber\\
& &\left.\sum_{x\in\calX}P(x)
\ln\left(\sum_{\delta}Q(\delta|x)e^{s\delta}\right)\right]\nonumber\\
&\ge&\min_{\{\calD:~\sum_{x\in\calX}P(x)\Delta_x\le
\Delta\}}\max_{s\le 0}
\left[s\Delta-\right.\nonumber\\
& &\left.\sum_{x\in\calX}P(x)
\ln\left(\sum_{\delta}Q(\delta|x)e^{s\delta}\right)\right]\nonumber\\
&=&
\max_{s\le 0}
\left[s\Delta-\sum_{x\in\calX}P(x)\ln\left(\sum_{\delta}Q(\delta|x)e^{s\delta}\right)\right]\nonumber\\
&=&I(\Delta),
\end{eqnarray}
where we have used the fact that the sum of maxima is cannot be smaller than
the maximum of a sum, as well as the fact that the optimum $s$ is to be sought
in the range $s\le 0$,
and so, $\sum_{x\in\calX}P(x)\Delta_x\le\Delta$ implies
$s\sum_{x\in\calX}P(x)\Delta_x\ge s\Delta$.

In the other direction, let $s^*$ be the achiever of $I(\Delta)$,
namely, the solution $s$ to the equation
$$\Delta=\frac{\partial}{\partial
s}\sum_{x\in\calX}P(x)\ln\left(\sum_{\delta}Q(\delta|x)e^{s\delta}\right)$$
and consider the distortion allocation
$$\Delta_x^*=\left[\frac{\partial}{\partial s}\ln
\left(\sum_{\delta}Q(\delta|x)e^{s\delta}\right)\right]_{s=s^*}$$
which obviously complies with the overall distortion constraint.
Thus,
\begin{eqnarray}
\tilde{I}(\Delta)&=&\min_{\{\calD:~\sum_{x\in\calX}P(x)\Delta_x\le
\Delta\}}\sum_{x\in\calX}P(x)\times\nonumber\\
& &\max_{s_x\le 0}\left[s_x\Delta_x-
\ln\left(\sum_{\delta}Q(\delta|x)e^{s_x\delta}\right)\right]\nonumber\\
&\le&
\sum_{x\in\calX}P(x)\cdot\max_{s_x\le 0}\left[s_x\Delta_x^*-
\ln\left(\sum_{\delta}Q(\delta|x)e^{s_x\delta}\right)\right]\nonumber\\
&=&
\sum_{x\in\calX}P(x)\left[s^*\Delta_x^*-
\ln\left(\sum_{\delta}Q(\delta|x)e^{s^*\delta}\right)\right]\nonumber\\
&=&s^*\Delta-
\sum_{x\in\calX}P(x)\ln\left(\sum_{\delta}Q(\delta|x)e^{s^*\delta}\right)\nonumber\\
&=&I(\Delta).
\end{eqnarray}
This completes the proof that $\tilde{I}(\Delta)=I(\Delta)$. $\Box$\\

\noindent
{\it Comment:} As noted in
\cite{Merhav08}, our discussion in this section,
as well as in the next section, applies to channel capacity
too, provided that $P=\{P(x)\}$ is understood as the channel output
distribution, $Q=\{Q(\hx)\}$ is the random (channel) coding distribution, the
distortion measure is taken to be $d(x,\hx)=-\ln W(x|\hx)$, where $W$ is
the transition probability matrix associated with the memoryless channel,
and the ``distortion level'' is set to
$\Delta=-\sum_{x,\hx}Q(\hx)W(x|\hx)\ln W(x|\hx)$. In this case, the maximizing $s$
is always $s^*=1$.

\section{Integral Representations}

In view of the observations made in Section 3,
it is interesting to represent the rate--distortion function as mechanical work
carried out on the distortion variable
along a reversible process, as well as in terms of the integrated variance of the distortion:
\begin{eqnarray}
R_Q(\Delta)&=&\sum_{x\in\calX}P(x)\cdot
\int_{\left<\delta\right>_{0|x}}^{\left<\delta\right>_{s|x}}\hs\cdot\mbox{d}
\left<\delta\right>_{\hs|x}\nonumber\\
&=&\sum_{x\in\calX}P(x)\cdot\int_0^s\mbox{d}\hs\cdot\hs\cdot\mbox{Var}_{\hs|x}\{\delta\},
\end{eqnarray}
where $s$ is related to $\Delta$ via the relation
$$ \sum_{x\in\calX}P(x)\left<\delta\right>_{s|x}=\Delta$$
and where $\left<\delta\right>_{s|x}$ and $\mbox{Var}_{s|x}\{\delta\}$ are
defined in the spirit of the earlier definitions of $\left<y\right>_s$ and
$\mbox{Var}_s\{y\}$ except that $y$ is replaced by $\delta$ and $P_s$ now
includes conditioning on $x$. I.e.,
$$\left<\delta\right>_{s|x}=\frac{\sum_\delta\delta
Q(\delta|x)e^{s\delta}}{\sum_\delta Q(\delta|x)e^{s\delta}}$$
and
\begin{eqnarray}
\mbox{Var}_{s|x}\{\delta\}
&=&\frac{\sum_\delta(\delta-\left<\delta\right>_{s|x})^2
Q(\delta|x)e^{s\delta}}{\sum_\delta Q(\delta|x)e^{s\delta}}\nonumber\\
&=&
\frac{\sum_\delta\delta^2
Q(\delta|x)e^{s\delta}}{\sum_\delta
Q(\delta|x)e^{s\delta}}-\left<\delta\right>_{s|x}^2.
\end{eqnarray}
Upper and lower bounds can be obtained from
\begin{eqnarray}
&&\sum_{x\in\calX}P(x)\cdot\sum_{i=1}^{\ell-1}s_i(\left<\delta\right>_{s_{i+1}|x}-\left<\delta\right>_{s_i|x})
\nonumber\\
&\le&
R_Q(\Delta)\nonumber\\
&\le&
\sum_{x\in\calX}P(x)\cdot\sum_{i=1}^{\ell-1}s_{i+1}(\left<\delta\right>_{s_{i+1}|x}-\left<\delta\right>_{s_i|x}).
\end{eqnarray}
The integrated variance formula above can also be represented as
$$R_Q(\Delta_s)=\int_0^s\mbox{d}\hs\cdot\hs\cdot\sum_{x\in\calX}P(x)\cdot\mbox{Var}_{\hs|x}\{\delta\}
=\int_0^s\mbox{d}\hs\cdot\hs\cdot\mbox{mmse}(\hs),$$
where $\mbox{mmse}(s)$ is the minimum mean squared error (MMSE)
in estimating the RV $\delta$ based on $x$, 
when they are jointly
distributed according to $P_s(x,\delta)=P(x)P_s(\delta|x)$,
with $P_s(\delta|x)$ being defined as
$$P_s(\delta|x)=\frac{Q(\delta|x)e^{s\delta}}{\sum_{\delta'}Q(\delta'|x)e^{s\delta'}}.$$
At the same time, the distortion itself, $\left<\delta\right>_s$, which we
also denote by $\Delta$,
can be represented using similar
integrals, but without the factor $\hs$ at the integrand:
\begin{eqnarray}
\Delta&\equiv&\left<\delta\right>_s\nonumber\\
&=&\sum_{x\in\calX}P(x)\cdot\left[\left<\delta\right>_{0|x}+
\int_0^s\mbox{d}\hs\cdot\mbox{Var}_{\hs|x}\{\delta\}\right]\nonumber\\
&=&\Delta_0+
\int_0^s\mbox{d}\hs\cdot\mbox{mmse}(\hs).
\end{eqnarray}

\noindent
{\it Example~3.}
Consider the binary symmetric source (BSS) and the Hamming distortion
measure. In this case, the optimum $Q$ is also symmetric. Here $\delta$
is a binary RV with $\mbox{Pr}\{\delta=1|x\}=e^s/(1+e^s)$ independently of 
$x$. Thus, the MMSE estimator of $\delta$ based on $x$ is 
$$\hat{\delta}=\frac{e^s}{1+e^s},$$ 
regardless of $x$,
and so the resulting MMSE is easily found to
be 
$$\mbox{mmse}(s)=\frac{e^s}{(1+e^s)^2}.$$
Accordingly,
$$\Delta=\frac{1}{2}+\int_0^s\frac{e^{\hs}\mbox{d}
\hs}{(1+e^{\hs})^2}=\frac{e^s}{1+e^s}$$
and
\begin{eqnarray}
R(\Delta)&=&\int_0^s\frac{\hs e^{\hs}
\mbox{d}\hs}{(1+e^{\hs})^2}\nonumber\\
&=&\ln 2 + \frac{se^s}{1+e^s}-\ln(1+e^s)\nonumber\\
&=&\ln 2-h_2\left(\frac{e^s}{1+e^s}\right)\nonumber\\
&=&\ln 2-h_2(\Delta),
\end{eqnarray}
where $h_2(u)=-u\ln u-(1-u)\ln(1-u)$ is the binary entropy function.
This concludes Example~3. $\Box$

The integrated 
variance expression can be generalized as follows: Let $\theta=t(x,\hx)$ be a given
function of $x$ and $\hx$ and let $\left<\theta\right>_s$ denote the
expectation of $t(x,\hx)$ w.r.t.\ the joint distribution of $x$ and
$\hx$ defined by
$$P_s(x,\hx)=\frac{P(x)Q(\hx)e^{sd(x,\hx)}}{\sum_{\hx'}Q(\hx')e^{sd(x,\hx')}}.$$
This characterizes the expected (and typical) value of
$\frac{1}{n}\sum_{i=1}^nt(x_i,\hx_i)$, where $\hx=(\hx_1,\ldots,\hx_n)$
continues to be the codeword that encodes $\bx$ from a rate--distortion code
designed and operated with the metric $d$.\footnote{As motivating examples, consider the
case where $t$ is another distortion measure -- although the codebook is
designed and operated relative to the metric $d$, its performance can also be
judged relative to an additional metric $t$. If $t(x,\hx)$ depends
on $\hx$ only, it may serve as a transmission power function $\Pi(\hx)$ (in joint source--channel
coding) or it can be the length function $\ell(\hx)$ (in bits) of lossless compression 
for the individual reproduction symbols.}
Then,
$$\left<\theta\right>_s=\left<\theta\right>_0+\int_0^s\mbox{d}\hs\cdot
\sum_{x\in\calX}P(x)\cdot\mbox{Cov}_{s|x}\{\theta,\delta\},$$
where $\mbox{Cov}_{s|x}\{\theta,\delta\}$ is the covariance between
$\theta=t(x,\hx)$ and $\delta=d(x,\hx)$, induced by 
$$Q_s(\hx|x)=\frac{Q(\hx)e^{sd(x,\hx)}}{\sum_{\hx'}Q(\hx')e^{sd(x,\hx')}},$$
for fixed $x$. 
This is integral form is a somewhat more general version 
of the fluctuation--dissipation theorem, mentioned above.

\section{Summary and Conclusion}

In this work, we have proposed another look at large deviations rate functions (or Chernoff
functions), where the Chernoff parameter is viewed as `force' rather
than as temperature. This leads to the interpretation of fundamental
quantities in information theory, like the rate--distortion function and channel capacity,
as free energies of certain physical systems. This interpretation has the
following advantages relative to the one proposed in \cite{Merhav08}:\\

\noindent
1) As explained in Subsection 2B, there is no need 
to interpret random coding distributions as degeneracy.\\

\noindent
2) As a consequence of 1),
we are able to construct an example of a physical system
whose behavior is analogous to that of
the rate--distortion coding problem. The properties of this system
were described in the second to the last paragraph of the Introduction.\\

\noindent
3) This interpretation generalizes to rate functions of combinations of rare
events. In this case, the rate function involves several Chernoff variables
(one per each event), which may correspond to a system with several forces,
each one acting on its own variable (cf.\
$R(\Delta_1,\Delta_2)$ in Subsection 2B). 
Our earlier physical example of a
one--dimensional array can now be extended to two dimensions, where the
elements are arranged in a rectangular 
lattice, and each
element has both a length and a width associated with each state. The sum
$[s_1\sum_id_1(x_i,\hx_i)+s_2\sum_id_2(x_i,\hx_i)]$ can be viewed as the inner product between a two
dimensional force vector and a two--dimensional displacement vector.
Alternatively, $s_1$ and $s_2$ may designate two different types of forces
(e.g., a mechanical force and a magnetic force). Either way, 
our derivations extend quite
straightforwardly to this setting.\\

\noindent
4) As mentioned before, we assumed throughout the derivation that
the random coding distribution is fixed, independently of the distortion
level, that is, independently of $s$. This is why we described $R(\Delta)$
as a process along the curve $R_Q(\cdot)$ with the understanding that $Q$
is chosen to be optimum for the target distortion $\Delta$. One can modify
the analysis to correspond to a process along $R(\cdot)$.
As mentioned earlier, however, in most cases, the optimum $Q$
depends on $s$,
and this dependency requires correction terms that
depend on the expected values of some derivatives of $\ln Q(\hx)$
w.r.t.\ $s$.
In the analogous physical 
interpretation proposed here, $s$ continues to be an
external control parameter that affects the Hamiltonian. The dependence
of the Hamiltonian on $s$ would now be non--linear, but this
may still be physically relevant.\\

\noindent
5) This interpretation as free energy opens the door to new points of view on the
rate--distortion function, e.g., as work done on the distortion variable
along a slow process, or as integrated variance (or MMSE).

\end{document}

%% file: rd.pstex_t
\begin{picture}(0,0)%
\includegraphics{rd.pstex}%
\end{picture}%
\setlength{\unitlength}{3947sp}%
\begingroup\makeatletter\ifx\SetFigFont\undefined%
\gdef\SetFigFont#1#2#3#4#5{%
  \reset@font\fontsize{#1}{#2pt}%
  \fontfamily{#3}\fontseries{#4}\fontshape{#5}%
  \selectfont}%
\fi\endgroup%
\begin{picture}(3628,3271)(239,-2584)
\put(2181,-2058){\makebox(0,0)[lb]{\smash{{\SetFigFont{7}{8.4}{\rmdefault}{\mddefault}{\itdefault}{$\lambda=kTR'(\Delta)$}%
}}}}
\put(1506,-2316){\makebox(0,0)[lb]{\smash{{\SetFigFont{6}{7.2}{\rmdefault}{\mddefault}{\itdefault}{$n\Delta_0$}%
}}}}
\put(926,-2026){\makebox(0,0)[lb]{\smash{{\SetFigFont{6}{7.2}{\rmdefault}{\mddefault}{\itdefault}{$n\Delta$}%
}}}}
\put(381,-1355){\makebox(0,0)[lb]{\smash{{\SetFigFont{6}{7.2}{\rmdefault}{\mddefault}{\itdefault}{$W=nkTR(\Delta)$}%
}}}}
\end{picture}%

%% file: chain3.pstex_t
\begin{picture}(0,0)%
\includegraphics{chain3.pstex}%
\end{picture}%
\setlength{\unitlength}{3947sp}%
\begingroup\makeatletter\ifx\SetFigFont\undefined%
\gdef\SetFigFont#1#2#3#4#5{%
  \reset@font\fontsize{#1}{#2pt}%
  \fontfamily{#3}\fontseries{#4}\fontshape{#5}%
  \selectfont}%
\fi\endgroup%
\begin{picture}(4224,678)(139,-252)
\put(1263,298){\makebox(0,0)[lb]{\smash{{\SetFigFont{9}{10.8}{\rmdefault}{\mddefault}{\itdefault}{$y_B$}%
}}}}
\put(2013,-91){\makebox(0,0)[lb]{\smash{{\SetFigFont{9}{10.8}{\rmdefault}{\mddefault}{\itdefault}{$L=nY$}%
}}}}
\put(4002,174){\makebox(0,0)[lb]{\smash{{\SetFigFont{9}{10.8}{\rmdefault}{\mddefault}{\itdefault}{$\lambda$}%
}}}}
\put(709,298){\makebox(0,0)[lb]{\smash{{\SetFigFont{9}{10.8}{\rmdefault}{\mddefault}{\itdefault}{$y_A$}%
}}}}
\end{picture}%